\documentclass[10pt]{article}

\usepackage{bm}
\usepackage{latexsym}
\usepackage{amssymb,amsmath,amsthm,amsfonts,url}

\newtheorem{defi}{Definition}
\newcommand{\ket}[1]{\vert#1\rangle}
\newcommand{\expec}[1]{\langle#1\rangle}
\newcommand{\A}{^{(A)}}
\newcommand{\B}{^{(B)}}

\begin{document}

\title{On local-hidden-variable no-go theorems}
\author{
Andr\'e Allan M\'ethot\\[0.5cm]
\normalsize\sl D\'epartement d'informatique et de recherche op\'erationnelle\\[-0.1cm]
\normalsize\sl Universit\'e de Montr\'eal, C.P.~6128, Succ.\ Centre-Ville\\[-0.1cm]
\normalsize\sl Montr\'eal (QC), H3C 3J7~~\textsc{Canada}\\
\normalsize\url{methotan@iro.umontreal.ca}
}

\maketitle
\begin{abstract}
The strongest attack against quantum mechanics came in 1935 in the
form of a paper by Einstein, Podolsky and Rosen. It was argued
that the theory of quantum mechanics could not be called a
complete theory of Nature, for every element of reality is not
represented in the formalism as such. The authors then put forth a
proposition: we must search for a theory where, upon knowing
everything about the system, including possible hidden variables,
one could make precise predictions concerning elements of reality.
This project was ultimatly doomed in 1964 with the work of Bell
Bell, who showed that the most general local hidden variable
theory could not reproduce correlations that arise in quantum
mechanics. There exist mainly three forms of no-go theorems for
local hidden variable theories. Although almost every physicist
knows the consequences of these no-go theorems, not every
physicist is aware of the distinctions between the three or even
their exact definitions. Thus we will discuss here the three
principal forms of no-go theorems for local hidden variable
theories of Nature. We will define Bell theorems, Bell
theorems without inequalities and pseudo-telepathy. A
discussion of the similarities and differences will follow.
\end{abstract}

\section{Introduction}\label{sec:intro}

In 1935, Einstein, Podolsky and Rosen (EPR) wrote their famous
paper ``Can quantum-mechanical description of physical reality be
considered complete?''~\cite{epr35}, wherein the authors answered
the question in the negative. Completeness was then considered an
important criterion regarding the validity of any physical theory
of Nature, and would still be for any proponent of determinism and
a classical view of the universe. Thus, their paper was a great
threat to the validity of quantum mechanics (QM).

Completeness is defined in the EPR paper as ``every element of the
physical reality must have a counterpart in the physical theory",
where physical reality should be interpreted as ``If, without in
any way disturbing a system, we can predict with certainty (i.e.,
with probability equal to unity) the value of a physical quantity,
then there exists an element of physical reality corresponding to
this physical quantity." EPR used the correlations obtained from
bipartite measurements of an entangled state to claim that
position and momentum can, do, and in fact must have simultaneous
realities. More precisely, they used a bipartite entangled state
and separated the two subsystems into space-like-separated
regions. In their particular case, if one was to perform a
position measurement on the first subsystem, from the result we
could determine the exact position of the second subsystem. The
same could be done for momentum. From their definition of physical
reality and since, according to relativity, the space-like
separation of the two subsystems prevents any interaction between
them, EPR concluded that position and momentum must have
simultaneous realities. From the fact the there are no
interactions between the subsystems, they also concluded that
these realities existed all along, from the time of the separation
of the subsystems. Since the formalism of QM precludes such a
description of reality, they were forced to conclude that QM
cannot be considered a complete theory of Nature. For EPR, Nature
possesses local hidden variables (LHVs), that can or cannot be
known, which determine the behavior of a system under any
measurement. In this picture, a measurement only reveals a
pre-existing property of Nature, while the usual consensus in
quantum mechanics is that a measurement forces a property of the
system into existence. Einstein then spent the rest of his life in
search of such a theory.

The LHV paradigm is the straightforward mathematical
representation of local realism. In a LHV model, the hidden
variables are set according to some probability distribution at
the creation of a state. They can only be accessed experimentally
though measurement, which perturbs the state and possibly alters
the hidden variables. If we were to know the actual values of the
hidden variables, we could fully predict the behaviour of the
system under any measurement of an element of reality, but this
lack of knowledge forces to average over the possible values of
the LHVs. Thus the Heisenberg uncertainty principle is not
violated and the probabilistic structure of QM is preserved. The
locality criterion means that the outcome of a measurement on
space-like separated systems cannot be correlated in any way other
than through the original hidden variables, which are fixed once
the state is created and change only according to local
operations.

Even though the EPR paper contained logical errors~\cite{bm05}, it
was still the firmest attack against a quantum description of the
physical world. Even the mightiest and hardiest defender of QM,
Bohr~\cite{bohr35}, was not able to firmly put the EPR argument to
rest~\cite{bm05}. It took the better part of three decades for a
complete refutation of the EPR argument to be put forth:
in~\cite{bell64}, Bell laid down the most general LHV model for a
particular measurement setup. He then showed that the expectation
value of the measurement operator could not, in \emph{any} LHV
model, come near the actual value predicted by QM. Experiments
have confirmed the correctness of the predictions of
QM~\cite{fc72,agr81}.

The work of Bell has been called ``the most profound discovery of
science"~\cite{stapp77}. At least, it could easily be argued that
Bell's theorem has changed our view of Nature in the same way as
Newton's classical mechanics and Einstein's relativity has. From
his work, we now know that conjugate operators do not have
simultaneous existence and only a measurement can force a state
which is not an eigenstate of the operator to bring a value to the
physical quantity attributed to the operator into existence.

In this paper, we present different forms of no-go theorems
concerning LHV theories. To the knowledge of the author, only
three forms of refutation exist, and a comprehensive comparative
study has not yet been published. It should be noted that this
paper does not have the goal of being an exhaustive survey of all
the LHV no-go theorems. In Section~\ref{sec:threeforms}, we will
give formal definitions to the three forms of no-go theorems,
namely Bell theorems, Bell theorems without
inequalities\footnote{Also called Bell inequalities without
probabilities, Bell inequalities without inequalities or all-versus-nothing refutation of EPR.} and
pseudo-telepathy. Section~\ref{sec:discussion} will then contain a
discussion of the similarities and differences of the three forms.


\section{The three forms}\label{sec:threeforms}

\subsection{Bell theorems}\label{sec:BI}

The first proof that the physical world could not be described by
any LHV theory came from Bell in 1964 in the form of an
inequality~\cite{bell64}. Bell bounded the absolute value of the
expectation value of a specific operator in \emph{any} LHV model
and showed that quantum mechanics violates this bound. As such, we
define a Bell theorem as follows.

\begin{defi}[Bell theorem]\label{def:bi}
A Bell theorem is a set of multipartite measurements on an
entangled state where the correlations obtained from the
measurements cannot be reproduced by any local classical model
where no communication between the participants is allowed.
\end{defi}

\subsubsection{An example of a Bell theorem}
The example we are to discuss here is often thought as the generic
Bell theorem and was put forth as an experimental proposal by
Clauser, Horne, Shimony and Holt~\cite{chsh69}. Suppose $A_1$ and
$A_2$ are measurement settings on Alice's apparatus and $B_1$ and
$B_2$ are measurement settings on Bob's. Given that the value of
the outcome to each measurement lies between $-1$ and 1, we can
easily bound the value of the operator $\expec{A_1 B_1} +
\expec{A_1 B_2} + \expec{A_2 B_1} - \expec{A_2 B_2}$ in any LHV
theory. Since the equation is linear in every variable and since
every variable must be independent of one another (locality
constraint) the maximum will be reached with every variable taking
an extremal value, 1 or $-1$. Therefore, we have $\vert \expec{A_1
B_1} + \expec{A_1 B_2} + \expec{A_2 B_1} - \expec{A_2 B_2} \vert
\leq 2$. In quantum mechanics, we can find appropriate
measurements on a singlet state, $(\ket{+-}-\ket{-+})/\sqrt{2}$,
to yield $\vert \expec{A_1 B_1} + \expec{A_1 B_2} + \expec{A_2
B_1} - \expec{A_2 B_2} \vert = 2\sqrt{2}$. The details are left as
an exercise to the reader. From the fact that no LHV model can
obtain a higher expectation value than 2, while QM can, is a clear
proof that quantum correlations cannot be simulated by a LHV
theory. If we now take the predictions of QM to be correct, we
have to forsake the search for a local realistic description of
reality.


\subsection{Bell theorems without inequalities}\label{sec:BIWI}

A Bell theorem appeals to a statistical argument and, as such,
does not look attractive to many. Thus, a more direct rebuttal of
LHVs was sought for. The first such proof came from Heywood and
Redhead~\cite{hr83}, where the authors aimed to propose an
experimental verification of the Kochen-Specker
theorem~\cite{ks67} to reject locality\footnote{The first example
is often wrongly attributed to Greenberger, Horne and
Zeilinger~\cite{ghz89}.}. It is important to note that Bell
inequalities without inequalities are not experimental proposals
which would rule out any LHV model in only one run. As Asher Peres
once said~\cite{peres}: ``The list of authors [who has made this
mistake] is too long to give explicitly, and it would be unfair to
give only a partial list." We will discuss this in more detail in
Section~\ref{sec:similarities}.

\begin{defi}[Bell theorems without
inequalities]\label{def:biwi} A Bell theorem without
inequalities is a set of multipartite measurements on an entangled
state where any local classical model, which is to attempt to
simulate the probability distribution of the outputs given by
quantum mechanics, will attribute a non zero probability to a
measurement outcome that is forbidden by quantum mechanics or will
never produce certain outcomes which are predicted with a non zero
probability in quantum mechanics.
\end{defi}

\subsubsection{An example of a Bell theorem without inequalities}
\label{sec:hardy}

For this example, we will give Brassard's~\cite{gilles} rendition
of Hardy's proof~\cite{hardy92}. Let us start with the state
$(\ket{--} + \ket{-+} + \ket{+-})/\sqrt{3}$ along the $z$ axis.
Let say that Alice and Bob are now given the choice of performing
either the $\sigma_z$ or $\sigma_x$ measurement. According to QM,
if Alice and Bob are to measure $\sigma_x\otimes\sigma_x$, then
they will receive the output $--$ with probability $1/12$. Let us
now assume that the state has LHVs that will produce a $--$ output
on a $\sigma_x\otimes\sigma_x$ measurement. From the criteria of
locality and realism, we now have that any local $\sigma_x$
measurement on this particular state will produce the output $-$.
Let us now see what happens if Alice and Bob are to measure
$\sigma_x\otimes\sigma_z$ or $\sigma_z\otimes\sigma_x$. From the
predictions of QM, the state should never be allowed to produce
$--$ as output. Once again according to the criteria of locality
and realism, we are forced to conclude that upon a local
$\sigma_z$ measurement, the state will produce $+$ as output.
Therefore, a $\sigma_z\otimes\sigma_z$ measurement on this
particular instance is bound to output $++$, which is forbidden by
QM. So in order for the LHV theory to output $--$ on a
$\sigma_x\otimes\sigma_x$ measurement with a non zero probability,
it will also output $++$ on a $\sigma_z\otimes\sigma_z$
measurement with non zero probability.


\subsection{Pseudo-telepathy}\label{sec:PT}

Pseudo-telepathy was first defined, although not yet termed as
such, in a paper by Brassard, Cleve and Tapp, where they turned
the Deutsch-Jozsa algorithm into a distributed form to show that
an exponential number of bits of communication was needed to
simulate the correlations of a certain number of singlet
states~\cite{bct99}. For a complete survey, please refer
to~\cite{bbt04a}.

\begin{defi}[Game]
A \emph{bipartite game} $G=(X,Y,R)$ is a set of inputs
$X=X\A\times X\B$, a set of outputs \mbox{$Y=Y\A\times Y\B$} and a
relation $R\subseteq X\A\times X\B\times Y\A\times Y\B$.
\end{defi}

\begin{defi}[Winning Strategy]
A \emph{winning strategy} for a bipartite game  $G=(X,Y,R)$ is a
strategy according to which for every $x\A\in X\A$ and $x\B\in X\B$,
Alice and Bob output $y\A$ and $y\B$ respectively such that
$(x\A,x\B,y\A,y\B)\in R$.
\end{defi}

\begin{defi}[Pseudo-telepathy]\label{def:pseudo-telepathy}
We say that a bipartite game $G$ exhibits \emph{pseudo-telepathy}
if  bipartite measurements of an entangled quantum state can yield
a winning strategy, whereas no classical strategy that does not
involve communication is a winning strategy.
\end{defi}

The extension to the multipartite case is trivial. The
generalization of Definition~\ref{def:pseudo-telepathy} can be
translated into a set of multipartite measurements on an entangled
state where any local classical model which is to attempt to
produce outputs that are not forbidden by quantum mechanics will
fail. We refer to this phenomenon by the term pseudo-telepathy
because of its shortness and its illustrativeness. To someone who
has no knowledge of quantum mechanics, and thus still believes in
local realism, these correlations between the measurement outcomes
can only be explained by some communication between the
sub-systems. Thus if we make these measurements in space-like
separated regions the sub-systems act as if they could
telepathically communicate instantaneously.

\subsubsection{An example of a pseudo-telepathy game}\label{sec:msg}

We present here the pseudo-telepathy game generally known as the
Magic Square game~\cite{aravind02}. The participants, namely Alice
and Bob, are each presented with a question: a random trit
$x\A\in\{0,1,2\}$ and $x\B\in\{0,1,2\}$ respectively. They must
produce three bits each, $y_1\A$, $y_2\A$, $y_3\A$ and $y_1\B$,
$y_2\B$, $y_3\B$ respectively. In order for them to win, $y_1\A +
y_2\A + y_3\A$ must be even, $y_1\B + y_2\B + y_3\B$ must be odd
and $y_{x\B}\A$ must equal $y_{x\A}\B$. In order for them to have
a winning strategy without resorting to QM or communicating, Alice
and Bob must share a $3\times 3$ table of 0s and 1s such that the
sum of the elements in each row is even and the sum of the
elements in each column is odd. A simple parity argument shows
that such a table is impossible. On the other hand, if Alice and
Bob are allowed to share entanglement, they can find a strategy
which, while convoluted, can be explicitely constructed. It should
be noted that they cannot construct the $3\times 3$ table required
classically, but they can use the correlations of the quantum
state to give three bits each respecting the above conditions.


\section{Discussion of the similarities and
differences}\label{sec:discussion}

\subsection{Similarities}\label{sec:similarities}

The obvious similarity between the three forms of no-go theorems
is that they all reject a local realistic description of Nature.
Therefore, the physical world cannot be described by any LHV
theory. As mentioned in Section~\ref{sec:intro}, all three require
many runs of the same experiment, with settings chosen at random
on each run, to rule out any LHV model. A LHV might be lucky and
answer correctly for many runs. Since a LHV theory can only
succeed with a marginal probability of success, but still can
succeed, we can only collect overwhelming evidence against LHVs.
The only rejection we can make in one run is of quantum mechanics
itself. If we consider perfect apparatus, then the production of a
forbidden output by quantum mechanics, in a Bell theorem
without inequalities or pseudo-telepathy experiment, would
invalidate the correctness of quantum mechanical predictions.

It is also interesting to remark that every pseudo-telepathy game
is a Bell theorem without inequalities and every Bell
inequality without inequalities is a Bell theorem. These facts
follows from Definition~\ref{def:bi}, \ref{def:biwi} and
\ref{def:pseudo-telepathy}. In pseudo-telepathy, the fact that no
LHV strategy can always output within the relation $R$ can be seen
as giving a non zero probability to an output that is forbidden in
quantum mechanics---quantum mechanics being able to never give
this output. And from Definition~\ref{def:biwi} and
Definition~\ref{def:pseudo-telepathy}, we clearly have a set of
measurements on an entangled state where the correlations obtained
from the measurements cannot be reproduced by any LHV model, hence
falling into Definition~\ref{def:bi}.


\subsection{Differences}\label{sec:differences}

Bell theorems and pseudo-telepathy are
in a sense only quantitatively different. Pseudo-telepathy is simply an inequality where the quantum
violation of the inequality reaches the maximal algebraic value.
We saw in Section~\ref{sec:msg} an example where we appeared not
to be concerned with reaching the maximal value of an inequality,
but even this example can be converted as an
inequality. The exercise is left to the reader.

It might be tempting to think that Bell theorems without
inequalities are the same as pseudo-telepathy. In fact, the list
of people who have made this error might also be too long to
include here. The reason is that in both of these paradigms, we
are concerned with proving a no-go theorem by a contradiction.
However, it was proven recently that no pseudo-telepathy game can
exist where the participants share only on entangled qubit, a
$2\times 2$ system~\cite{bmt05}. For pseudo-telepathic
correlations to arise we need at least a $3\times 3$
system~\cite{hr83} or a $2\times 2\times 2$
system~\cite{mermin90a}. Therefore, Hardy's state cannot yield
correlations strong enough for pseudo-telepathy. As a consequence,
pseudo-telepathy and Bell theorems without inequalities are
different paradigms. The difference is buried subtly in
Definition~\ref{def:biwi} and
Definition~\ref{def:pseudo-telepathy}. In the first case, we
require the LHV model to be able to generate all the possible
outputs according to QM without outputting a forbidden output,
while in the second case the requirement is weaker; we only need
to produce outputs that can be generated by QM. In other words, in
pseudo-telepathy, we only need to avoid forbidden outputs, but we
do not need to produce every possible output. Hardy's proof relies
on the fact that once in a while the experiment will generate $++$
on a $\sigma_x\otimes\sigma_x$ measurement, but for a
pseudo-telepathy game it is of no help.


\section{Conclusion}\label{sec:conclusion}

We have seen the formal definition of the three forms of LHV no-go
theorems and how they differ one from another. We have thus shown
that it is important to look closely at our models when we want to
describe nature, for we could have seen that there is a
qualitative difference between Hardy's no-go theorems and most of
the other Bell theorems without inequalities. In this
hierarchy of no-go theorems, pseudo-telepathy is the stronger
refutation of the local realistic viewpoint as it lies at the top
of the LHV no-go theorem hierarchy. Not every Bell theorem is a
Bell theorem without inequalities and not every Bell theorem
without inequalities is a pseudo-telepathy game, while the
converse is true. There exists states that can generate
correlations strong enough for a Bell theorem without
inequalities while the same state cannot yield a pseudo-telepathy
game~\cite{bmt05}.


\section*{Acknowledgements}

The author would like to thank Gilles Brassard for many invaluable
discussions on the subject and Anne Broadbent and Richard
MacKenzie for very helpful comments.


\end{document}